\documentclass[citeautoscript,aps,pra,superscriptaddress,cite,amsmath,amsfonts,amssymb,twocolumn]{revtex4}
\renewcommand{\vec}[1]{\mathbf{#1}}
\usepackage{graphicx}
\usepackage{epsfig}
\usepackage{color}
\usepackage{url}

\def\urltilda{\kern -.15em\lower .7ex\hbox{\~{}}\kern .04em}

\renewcommand{\Re}{\operatorname{Re}}
\renewcommand{\Im}{\operatorname{Im}}

\newcommand{\eqnumref}[1]{(\ref{eq:#1})}
\renewcommand{\eqref}[1]{Eq.~\eqnumref{#1}}

\newcommand{\citeasnoun}[1]{Ref.~\onlinecite{#1}}

\begin{document}
\title{Calculation of nonzero-temperature Casimir forces in the time domain}

\author{Kai Pan}
\affiliation{Department of Physics,
Massachusetts Institute of Technology, Cambridge, MA 02139, USA}
\affiliation{Research Laboratory of Electronics,
              Massachusetts Institute of Technology,
              Cambridge, MA 02139, USA}
\author{Alexander~P.~McCauley}
\affiliation{Department of Physics,
Massachusetts Institute of Technology, Cambridge, MA 02139, USA}
\author{Alejandro~W.~Rodriguez}
\affiliation{Department of Physics,
Massachusetts Institute of Technology, Cambridge, MA 02139, USA}
\author{M.~T.~Homer Reid}
\affiliation{Department of Physics,
Massachusetts Institute of Technology, Cambridge, MA 02139, USA}
\affiliation{Research Laboratory of Electronics,
              Massachusetts Institute of Technology,
              Cambridge, MA 02139, USA}
\author{Jacob~K.~White}
\affiliation{Research Laboratory of Electronics,
              Massachusetts Institute of Technology,
              Cambridge, MA 02139, USA}
\affiliation{Department of Electrical Engineering and Computer Science,
              Massachusetts Institute of Technology,
              Cambridge, MA 02139, USA}

\author{Steven~G.~Johnson}
\affiliation{Research Laboratory of Electronics,
              Massachusetts Institute of Technology,
              Cambridge, MA 02139, USA}
\affiliation{Department of Mathematics,
Massachusetts Institute of Technology, Cambridge, MA 02139, USA}

\date{\today}

\begin{abstract}
We show how to compute Casimir forces at nonzero temperatures with time-domain electromagnetic simulations, for example using a finite-difference time-domain (FDTD) method. Compared to our previous zero-temperature time-domain method, only a small modification is required, but we explain that some care is required to properly capture the zero-frequency contribution. We validate the method against analytical and numerical frequency-domain calculations, and show a surprising high-temperature disappearance of a non-monotonic behavior previously demonstrated in a piston-like geometry.
\end{abstract}

\maketitle

In this paper, we show how to compute nonzero-temperature ($T > 0$)  corrections to
Casimir forces via time-domain calculations,  generalizing a computational approach based
on the finite-difference  time-domain (FDTD) method that we previously demonstrated for
$T=0$ ~\cite{RodriguezMc09:PRA,application}. New computational methods for Casimir interactions have
become important in order to model non-planar  micromechanical systems where unusual
Casimir effects have been  predicted ~\cite{Antezza06,RodriguezMc09:PRA,application,
  Rodriguez07:PRL, Rodriguez08:PRL, Emig07, Emig07:ratchet,
  Rodriguez07:PRA, ReidRo09,Pasquali08,Pasquali09,gies06:edge}, and there has been increasing interest in $T > 0$  corrections ~\cite{milton04,milton06, lamoreaux08, Alejdro:temp, paris:temp, germany:temp1, germany:temp2}, especially in recently identified systems where these effects are non-negligible ~\cite{Alejdro:temp}.  Although $T > 0$ effects are easy  to incorporate in the imaginary frequency domain, where they merely  turn an integral into a sum over Matsubara frequencies ~\cite{Landau:stat2}, they  turn out to be nontrivial to handle in the time-domain because of the  singularity of the zero-frequency contribution, and we show that a  naive approach leads to incorrect results. We validate our approach  both with a
one-dimensional system where analytical solutions are  available, and also in a two-dimensional (2D) piston-like geometry ~\cite{Rodriguez07:PRL,Rodriguez07:PRA,nonmonotonic,Zaheer} where  we compare to a frequency-domain numerical method. In the 2D piston  geometry, we observe an interesting effect in which a non-monotonic phenomenon previously identified at $T=0$ disappears for a  sufficiently large $T$.

The Casimir force is a combination of fluctuations
at all frequencies $\omega$, and the $T=0$ force can be expressed as an
integral $F(0) = \int_0^\infty f(\xi) d\xi$ over Wick-rotated imaginary frequencies
$\omega = i\xi$ ~\cite{Landau:stat2}. At a nonzero $T$, this integral is replaced by
a sum over ``Matsubara frequencies'' $\xi_{n}= n \pi \omega_T$ for
integers $n$, where $\omega_T=2 k_{\mathrm{B}}T/\hbar$ and $k_{\mathrm{B}}$ is Boltzmann's constant ~\cite{Landau:stat2}:
\begin{equation}
  F(T) = \pi \omega_T \left[ \frac{f(0^+)}{2} +
    \sum_{n=1}^\infty f\left(n \pi \omega_T \right) \right].
\label{eq:F-T}
\end{equation}
The transformation from the $T=0$ integral to a summation can be derived directly by considering thermodynamics in the Matsubara formalism. Eq.~(\ref{eq:F-T}) corresponds to a trapezoidal-rule approximation of the $T=0$ integral ~\cite{Steven:book}. At room temperature, the Matsubara frequency corresponds to a wavelength 2$\pi/\xi = 7\, \mu \textrm{m}$, much larger than separations where the Casimir effect is typically observed, so  usually $T>0$ corrections are negligible ~\cite{milton04,milton06}. However, experiments are pushing towards $>1\mu$m separations ~\cite{exp1,exp2,exp3} in an attempt to observe these corrections. Also, a recent theoretical  prediction shows much larger $T$ corrections with appropriate material and geometry choices ~\cite{Alejdro:temp}.

To compute Casimir forces in arbitrary geometries, it is desirable to exploit mature methods from
computational classical electromagnetism (EM), and a number of approaches have been suggested ~\cite{Taflove00,chew01,Jin02,boyd01:book,Hackbush89}. One technique is to use the fluctuation-dissipation theorem: the mean-square electric and  magnetic fields $\langle E^2 \rangle$ and $\langle H^2 \rangle$ can thereby be computed from classical Green's functions ~\cite{Landau:stat2}, and the mean stress tensor can be computed and integrated to obtain the force ~\cite{RodriguezMc09:PRA,application,Rodriguez07:PRA}. In particular, at each $\omega$, the correlation function of the fields $\langle E^2 \rangle$ is given by:
\begin{multline}
\langle E_{j}(\vec{x})E_{k}(\vec{x}')\rangle_{\omega}=\\
-\frac{\hbar}{\pi}\Im\left[\omega^{2}G_{jk}^{E}
(\omega;\vec{x},\vec{x}')\right]\coth\left(\frac{\omega}{\omega_T}\right),
\label{eq:Ecorr-real}
\end{multline}
where $G_{jk}^{E}=(\vec{G}_{k}^{E})_{j}$ is the classical dyadic
{}``photon'' Green's function, proportional to the electric field in the $j$ direction at $\vec{x}$ due to an electric-dipole current in the $k$ direction at $\vec{x}'$, and
solves
\begin{multline}
\left[\nabla\times\mu(\omega,\vec{x})^{-1}\nabla\times{}-\omega^{2}\varepsilon(\omega,\vec{x})
\right]\vec{G}_{k}^{E}(\omega,\vec{x},\vec{x}')\\
=\delta^{3}(\vec{x}-\vec{x}')\hat{e}_{k},
\label{eq:Green-real}
\end{multline}
where $\varepsilon$ is the electric permittivity tensor, $\mu$ is
the magnetic permeability tensor, and $\hat{e}_{k}$ is a unit vector
in direction $k$. The magnetic-field correlation $\langle H^2 \rangle$ has a similar form ~\cite{RodriguezMc09:PRA,application}. Note that the temperature dependence appears as a coth factor (from a Bose-Einstein distribution). If this is Wick-rotated to imaginary frequency $\omega=i\xi$, the poles in the coth function gives the sum ~(\ref{eq:F-T}) over Matsubara frequencies ~\cite{Lamoreaux05}. In our EM simulation, what is actually computed is the electric or magnetic field in response to an electric or magnetic dipole current, respectively. This is related to $G_{ij}$ by
\begin{equation}
E_{jk}(\omega,\textbf{x},\textbf{x}') = -i\omega G_{jk}^{E}(\omega,\textbf{x},\textbf{x}')
\label{eq:GE}
\end{equation}
where $E_{jk}(\omega,\textbf{x},\textbf{x}')$ denotes the electric field response in the $j$th direction
due to a dipole current source
$\textbf{J}(\omega,\textbf{x},\textbf{x}')=\delta(\textbf{x}-\textbf{x}')
\hat{\mathrm{e}}_k$ ~\cite{RodriguezMc09:PRA}.

This equation can be solved for each point on a surface to  integrate the stress
tensor, and for each frequency to integrate the  contributions of fluctuations at all
frequencies. Instead of computing each $\omega$ separately, one can use a pulse
source in time, whose Fourier transform contains all frequencies. As derived
in detail elsewhere \cite{RodriguezMc09:PRA,application}, it turns out that this corresponds to a sequence of time-domain simulations, where pulses of current are injected and some function $\Gamma(t)$
of the resulting fields  (corresponding to the stress tensor) is integrated in time,
multiplied  by an appropriate weighting factor $g(t)$.  We perform these simulations by using
the standard FDTD technique \cite{Taflove00}, which discretizes space and time on a uniform grid.
In frequency domain, Wick rotation to complex $\omega(\xi)$ is crucial for  numerical
computations in order to obtain a tractable frequency  integrand \cite{Rodriguez07:PRA,Steven:book}, and the analogue in time domain is equally important to obtain rapidly decaying fields (and hence short
simulations) \cite{RodriguezMc09:PRA,application}. In time domain, one must implement complex $\omega$ indirectly: because $\omega$ only
appears explicitly with $\varepsilon$ in Eq.~(\ref{eq:Green-real}), converting $\omega$ to the complex  contour $\omega(\xi) \equiv \xi \sqrt{1+\frac{i\sigma}{\xi}}$ is equivalent to operating at a real frequency $\xi$ with an artificial conductivity $\varepsilon(\vec{r}) \rightarrow \varepsilon(\vec{r}) (1 + \frac{i \sigma}{ \xi})$ ~\cite{RodriguezMc09:PRA,application}. One cannot use purely imaginary frequencies $\omega = i\xi$ in the time domain, because the corresponding material $\varepsilon \to -\varepsilon$ has exponentially growing solutions in time~\cite{RodriguezMc09:PRA}. Thus, by adding an artificial conductivity  everywhere, and including a corresponding Jacobian factor in $g(t)$, one obtains the same (physical) force result in a much shorter time (with  the
fields decaying exponentially due to the conductivity).

\begin{figure}[tb]
\includegraphics[width=1\columnwidth] {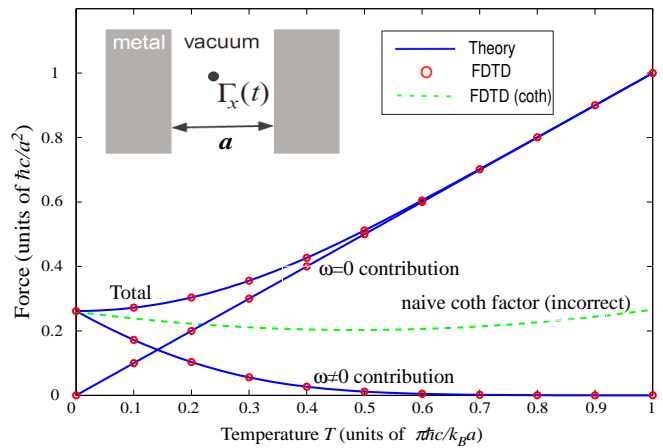}
\caption{Comparison between FDTD (red circles) and the analytical Lifshitz formula ~\cite{Lifshitz} (blue line) for the Casimir force between perfect-metal plates in 1D with separation $a$. The $\omega=0$ and $\omega \neq 0$ contributions to the Matsubara sum ~(\ref{eq:F-T}) are plotted separately, in addition to the total force. The straightforward method of including the $\coth(\hbar\omega/2kT)$ Boes-Einstein factor in the FDTD integration (green dashed line) gives an incorrect result because the $\omega=0$ pole requires special handling. }
\label{fig:parallel-plate}
\end{figure}

Now, we introduce the basic idea of how $T>0$ is incorporated in the time domain,  and explain where
the difficulty arises. The standard $T>0$ analysis of Eq.~(\ref{eq:F-T}) is expressed in the frequency domain, so we start there by exploiting the fact that the time-domain approach is derived from a Fourier transform of the frequency-domain
approach. In particular, $g(t)$ is the  Fourier transform of a weighting factor $g[\omega(\xi)]$
in the Fourier domain ~\cite{RodriguezMc09:PRA,application}. At real frequency, the effect of $T>0$ is simply to include an additional weighting factor $\coth\left[\frac{\omega(\xi)}{\omega_{T}}\right]$ in the $\omega(\xi)$ integral from Eq.~(\ref{eq:Ecorr-real}).  So, a straightforward, but naive, approach is to replace $g[\omega(\xi)]$ with:
\begin{multline}
g[\omega(\xi)] \rightarrow  g[\omega(\xi)] \coth\left[\frac{\omega(\xi)}{\omega_{T}}\right] \\
= -i\xi \left(\sqrt{1 + \frac{i \sigma}{\xi}}\right) (1+i\sigma / 2 \xi) \coth\left[\frac{\omega(\xi)}{\omega_{T}}\right],
\label{eq:gxicoth}
\end{multline}
using the $T=0$ $g\left[\omega(\xi)\right]$ expression from \citeasnoun{RodriguezMc09:PRA}, and then Fourier transform this to yield $g(t)$.  However, there is an  obvious problem
with this approach: the $1/\omega$ singularity in  $\coth\left[\frac{\omega(\xi)}{\omega_{T}}\right]$ means that Eq.~(\ref{eq:gxicoth}) is not locally integrable  around $\xi=0$, and therefore its Fourier transform is not well-defined. If we naively ignore this problem, and compute the Fourier transform via a discrete Fourier transform as in ~\cite{RodriguezMc09:PRA,application}, simply assigning an arbitrary finite value for the $\xi=0$ term, this unsurprisingly gives an incorrect
force for $T>0$ compared to the analytical Lifshitz formula for the case of parallel perfect-metal plates in 1D \cite{Lifshitz}, as shown in Fig.~\ref{fig:parallel-plate} (green dashed line).

Instead, a natural solution is to handle $\omega \neq 0$ by the coth factor as in Eq.~(\ref{eq:gxicoth}), but to subtract the $\omega=0$ pole and handle this contribution separately. As explained below, we will extract the correct $\omega=0$ contribution from the frequency-domain expression Eq.~(\ref{eq:F-T}), convert it to time domain, and add it back in as a manual correction to $g(t)$. In particular , the $\coth\left[\frac{\omega(\xi)}{\omega_{T}}\right]$ function has poles at $\omega = i n \pi \omega_{T}$ for integers $n$. When the frequency integral is Wick-rotated to imaginary frequency, the residues of these poles give the Matsubara sum Eq.~(\ref{eq:F-T}) via contour integration ~\cite{Lamoreaux05}. If we subtract the $n=0$ pole from the coth, obtaining
\begin{eqnarray}
{{g}_{n>0}}(\xi) = g[\omega(\xi)] \left\{ \coth \left[ \frac{\omega(\xi)}{\omega_T}\right]-\frac{\omega_T}{\omega(\xi)} \right\},
\label{eq:gxi}
\end{eqnarray}
the result of the time-domain integration of $g_{n>0}(t)\Gamma(t)$ will therefore correspond to all of the $n>0$ terms in Eq.~(\ref{eq:F-T}), nor is there any problem with the Fourier transformation to $g_{n>0}(t)$. Precisely this result is shown for the 1D parallel plates in Fig.~\ref{fig:parallel-plate}, and we see that it indeed matches the $n>0$ terms from the analytical expression. To handle the $\omega=0$ contribution, we begin with the real-$\omega$ $T=0$ expression for the Casimir force following our notation from the time-domain stress-tensor method \cite{RodriguezMc09:PRA,application}:
\begin{equation}
F_i = \Im \frac{\hbar}{\pi} \int_{0}^\infty d\omega ~g_R(\omega)
\Gamma_{i}(\omega),
\label{eq:Force_fields}
\end{equation}
where $g_R(\omega) = -i \omega$ is the weighting factor for the $\sigma=0$ real $\omega$ contour and $\Gamma_{i}(\omega)= \Gamma^E_i(\omega) + \Gamma^H_{i}(\omega)$ is the surface-integrated stress tensor (electric- and magnetic-field contributions). From Eq.~(\ref{eq:F-T}), the $\omega=0$ contribution for $T>0$ is then
\begin{eqnarray}
F_{i,(n=0)} &=& \lim_{\omega\rightarrow 0^+} \Im \left[\frac{\hbar}{\pi} \frac{1}{2} (-i\omega) \Gamma_i (\omega) \frac{2\pi k_{\mathrm{B}}T}{\hbar}\right] \\
&=& \lim_{\omega\rightarrow 0^+} \Re \left[-\omega \Gamma_i (\omega)  k_{\mathrm{B}}T\right].
\label{eq:zero}
\end{eqnarray}

Notice that $\hbar$ cancels in the $\omega=0$ contribution:
this term dominates in the limit of large $T$ where the fluctuations can be thought of as purely
classical thermal fluctuations. To relate Eq.~(\ref{eq:zero}) to what is actually computed in the
FDTD method requires some care because of the way in which we transform to the $\omega(\xi)$ contour. The quantity $\Gamma^E_i(\omega)$ is proportional to an integral of $E_{ij}(\omega)=-i \omega G_{ij}(\omega)$, from Eq.~(\ref{eq:GE}). However, the $\omega(\xi)$ transformed system computes $\tilde{\Gamma}^E_i(\xi) \sim \tilde{E}_{ij}(\xi)=-i \xi \tilde{G}_{ij}(\xi)$, where $\tilde{G}(\xi)$ solves Eq.~(\ref{eq:Green-real}) with $\omega^{2}\varepsilon(\vec{r}) \rightarrow \xi^2 (1 + \frac{i \sigma}{ \xi})\varepsilon(\vec{r})$, but what we actually want is $-i\omega G_{ij}(\omega)|_{\omega=\omega(\xi)}=-i \omega(\xi)\tilde{G}_{ij}(\xi)$. Therefore, the correct $\omega=0$ contribution is given by
\begin{equation}
\lim_{\omega\rightarrow 0^+} \Gamma^E_i(\omega)=\lim_{\xi\rightarrow 0^+} \frac{\omega(\xi)}{\xi} \tilde{\Gamma}^E_i(\xi)
\end{equation}
Combined with $\omega(\xi) k_{\mathrm{B}}T$ factor from Eq.~(\ref{eq:zero}), this gives an $n=0$ contribution of $\tilde{\Gamma}|_{\xi=0^+}$ multiplied by $-\omega(\xi)^2 k_{\mathrm{B}}T/\xi|_{\xi=0^+}=~\sigma k_{\mathrm{B}}T$. This $\omega=0$ term corresponds to a simple expression in the time domain, since $\tilde{\Gamma}|_{\xi=0^+}$ is simply the time integral of $\tilde{\Gamma}(t)$ and the coefficient $\sigma k_{\mathrm{B}}T$ is merely a constant. Therefore, while we originally integrated $g_{n>0}(t)\tilde\Gamma(t)$ to obtain the $n>0$ contributions, the $n=0$ contribution is included if we instead integrate:
\begin{equation}
\left[g_{n>0}(t) + \sigma k_{\mathrm{B}}T\right]\tilde{\Gamma}(t).
\label{eq:gt}
\end{equation}
The term $\left[g_{n>0}(t) + \sigma k_{\mathrm{B}}T\right]$ generalizes the original $g(t)$ function from ~\citeasnoun{RodriguezMc09:PRA} to any $T\geq0$.

We check Eq.~(\ref{eq:gt}) for the 1D parallel plate case in Fig.~\ref{fig:parallel-plate} against the analytical Lifshitz formula ~\cite{Lifshitz}. As noted above, the $g_{n>0}$ term ~(\ref{eq:gxi}) correctly gives the $n>0$ terms, and we also see that the $\sigma k_{\mathrm{B}}T$ term gives the correct $n=0$ contribution, and hence the total force is correct.

\begin{figure}[tb]
\includegraphics[width=1\columnwidth] {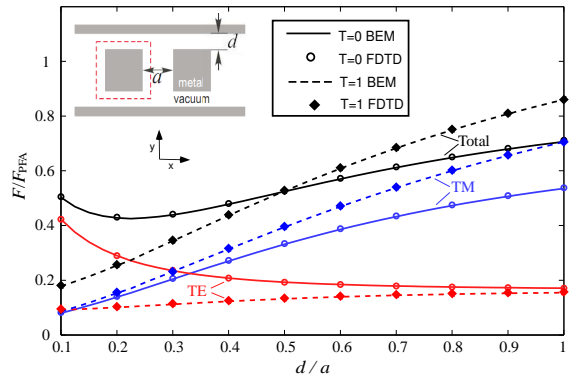}
\caption{Comparison between FDTD (circles and diamonds) and BEM frequency domain (solid and dashed lines) calculation of the 2D Casimir force ($z$-invariant fluctuations) between two perfect-metal sidewalls (separation $d$), normalized by the proximate force approximation for the 2D parallel plates $F_{PFA}=\hbar c \zeta(3)/8 \pi a^2$. At $T=0$ (circles and solid lines) total force (black) varies non-monotonically with $d$, due to competition between TE (red) and TM (blue) polarizations \cite{nonmonotonic}. At $T=1 \times \pi c \hbar/ k_{\mathrm{B}} a$ (dashed lines and diamonds) BEM and FDTD match, but the non-monotonicity disappears.}
\label{fig:combine}
\end{figure}

As another check, we consider a more complicated geometry: a piston-like configuration from ~\citeasnoun{Rodriguez07:PRL}, shown schematically in the inset of Fig.~\ref{fig:combine}. This system consists of two square rods adjacent between two sidewalls, which we solve here for the 2D case of $z$-invariant fluctuations. At $T=0$, such geometries were shown to exhibit an interesting non-monotonic variation of the force between the two blocks as a function of sidewall separation $d$ ~\cite{Rodriguez07:PRL,Rodriguez07:PRA,nonmonotonic,Zaheer}, which does not arise in the simple pairwise-interaction heuristic picture of the Casimir force. This can be seen in the solid lines of Fig.~\ref{fig:combine}, where the non-monotonicity arises from a competition between forces from transverse-electric (TE) and transverse-magnetic (TM) field polarizations ~\cite{Rodriguez07:PRL}, which in turn can be explained by a method-of-images argument ~\cite{nonmonotonic}. In Fig.~\ref{fig:combine}, the solid lines are computed by a $T=0$ frequency-domain boundary element method (BEM) evaluating a path-integral expression ~\cite{ReidRo09}, whereas the circles are computed by the $T=0$ FDTD method ~\cite{RodriguezMc09:PRA,application}, and both methods agree. We also compute the force at $T=1 \times \pi c \hbar/ k_{\mathrm{B}} a$ where the $\xi=0^+$ term dominates. We see that the FDTD method with the $T>0$ modification Eq.~(\ref{eq:gt}) (diamonds) agrees with the frequency-domain BEM results (dashed lines), where the latter simply use the Matsubara sum ~(\ref{eq:F-T}) to handle $T>0$.

Interestingly, Fig.~\ref{fig:combine} shows that the non-monotonic effect disappears for $T=1 \times \pi c \hbar/ k_{\mathrm{B}} a$, despite the fact that the method-of-images argument of ~\citeasnoun{nonmonotonic} ostensibly applies to the $\xi = 0^+$ quasi-static limit (which dominates at this large $T$) as well as to $\xi>0$. The argument used the fact that TM fluctuations can be described by a scalar field with Dirichlet boundary conditions (vanishing at the metal), and in this case the sidewalls introduce opposite-sign mirror sources that reduce the interaction as $d$ decreases; in contrast, TE corresponds to a Neumann scalar field (vanishing slope), which requires same-sign mirror sources that increase the interaction as $d$ decreases ~\cite{nonmonotonic}. In Fig.~{\ref{fig:combine}}, however, while the $T=1 \times \pi c \hbar/ k_{\mathrm{B}} a$ TM force still decreases as $d$ decreases, the TE force no longer increases for decreasing $d$ at $T=1 \times \pi c \hbar/ k_{\mathrm{B}} a$. The problem is that the image-source argument most directly applies to $z$-directed dipole sources in the scalar-field picture---electric $J_z^E$ currents for TM and magnetic $J_z^H$ currents for TE---while the situation for in-plane sources (corresponding to derivative of the scalar field from dipole-like sources) is more complicated~\cite{Emigadd}. For a sufficiently large $T$ dominated by the $\xi=0^+$ contribution (as is the case here), we find numerically that the $J_z^H$ sources as $\xi \rightarrow 0^+$
no longer contribute to the force. Intuitively, as $\xi \to 0^+$ a magnetic dipole source produces a more and more constant (long wavelength) field, which automatically satisfies the Neumann boundary conditions and hence is not affected by the geometry.  Instead, numerical calculations show that the TE $\xi = 0^+$ contribution is dominated by $J_x^E$ sources and the corresponding electric stress-tensor terms, which turn out to slightly \emph{decrease} in strength as $d$ decreases. (A related effect is that, for small $d$, it can be observed in Fig.~\ref{fig:combine} that the $T=1$
force is actually smaller than the $T=0$ force, again  due to the suppression of the TE
contribution. Since the force diverges as $T\to\infty$, this means that the force
changes non-monotonically with $T$ at small $d$; a similar non-monotonic temperature
dependence was previously observed for Dirichlet scalar-field fluctuations in a
sphere-plate geometry ~\cite{germany:temp2}.)

In contrast, if we consider the 3D constant cross-section problem with $z$-dependent fluctuations, corresponding to integrating $e^{i k_z z}$ fluctuations over $k_z$ ~\cite{Landau:stat2}, then we find that the non-monotonic effect is preserved at all $T$. This is easily explained by the fact that, for perfect metals, $k_z \neq 0$ is mathematically equivalent to a problem at $k_z = 0$ and $\xi \to \sqrt{\xi^2 +k_z^2}$ ~\cite{Rodriguez07:PRA, kz}, and so the $n=0$ Matsubara term still contains contributions equivalent to $\xi>0$ in which the $J_z^H$ mirror argument applies and the situation is similar to $T=0$. In any case, this 2D disappearance of non-monotonicity seems unlikely to be experimentally relevant, because we find that it only occurs for $T \gtrsim 0.7 \times \pi c \hbar/ k_{\mathrm{B}} a$ , which for $a=1\mu \textrm{m}$ corresponds to $T \gtrsim 5000K$.

The main point of this paper is that a simple (but not too simple) modification to our previous time-domain method allows off-the-shelf FDTD software to easily calculate Casimir forces at nonzero temperatures. Although the disappearance of non-monotonicity employed here as a test case appears unrealistic, recent predictions of other realistic geometry/material effects ~\cite{Alejdro:temp}, combined with the fact that temperature effects in complex geometries are almost unexplored at present, lead us to hope that future work will reveal further surprising temperature effects that are observable in micromechanical systems.

This work was supported in part by the Singapore-MIT Alliance Computational Engineering
Flagship program, by the Army Research Office through the ISN under Contract No. W911NF-07-D-0004, by US DOE Grant No. DE-FG02-97ER25308, and by the Defense Advanced Research Projects Agency (DARPA) under contract N66001-09-1-2070-DOD.


\end{document}